\documentclass[nonacm]{acmart}

\usepackage{makecell}
\usepackage{multirow}
\usepackage{xcolor,colortbl}
\usepackage{dirtytalk}
\usepackage{comment}

\AtBeginDocument{%
  \providecommand\BibTeX{{%
    \normalfont B\kern-0.5em{\scshape i\kern-0.25em b}\kern-0.8em\TeX}}}

\begin{document}

\title{Towards evaluating and eliciting high-quality documentation for intelligent systems}

\author{David Piorkowski}
\email{djp@ibm.com}
\affiliation{
  \institution{IBM Research}
  \city{Yorktown Heights}
  \state{New York}
  \postcode{10606}
  \country{United States}
}

\author{Daniel Gonz\'alez}
\email{gonzad5@rpi.edu}
\affiliation{
  \institution{Rensselaer Polytechnic Institute}
  \city{Troy}
  \state{New York}
  \postcode{12180}
  \country{United States}
}

\author{John Richards}
\email{ajtr@us.ibm.com}
\affiliation{
  \institution{IBM Research}
  \city{Yorktown Heights}
  \state{New York}
  \postcode{10606}
  \country{United States}
}

\author{Stephanie Houde}
\email{Stephanie.Houde@ibm.com}
\affiliation{
  \institution{IBM Research}
  \city{Cambridge}
  \state{Massachusetts}
  \postcode{02142}
  \country{United States}
}

\renewcommand{\shortauthors}{Piorkowski et al.}

\begin{abstract}
A vital component of trust and transparency in intelligent systems built on machine learning and artificial intelligence is the development of clear, understandable documentation. However, such systems are notorious for their complexity and opaqueness making quality documentation a non-trivial task. Furthermore, little is known about what makes such documentation "good." In this paper, we propose and evaluate a set of quality dimensions to identify in what ways this type of documentation falls short. Then, using those dimensions, we evaluate three different approaches for eliciting intelligent system documentation. We show how the dimensions identify shortcomings in such documentation and posit how such dimensions can be use to further enable users to provide documentation that is suitable to a given persona or use case. 
\end{abstract}

\maketitle
\section{Introduction}

As Artificial Intelligence (AI)\footnote{For the purposes of this paper, AI includes machine learning} models and services are used in a growing number of high-stakes applications, interest has grown around the need for a clear record of their purpose, appropriate use, technical structure, training data, accuracy, fairness, and robustness \cite{raji2019ml,arnold2019factsheets}. This is motivated by ethical and legal concerns as well as communication needs among technical as well as non-technical stakeholders involved in initiating, producing, deploying, monitoring, and ultimately consuming AI services \cite{brundage2020toward,havrda2020enhanced}.

Several proposals for high quality and transparent AI reporting frameworks have emerged \cite{mitchell2019model,gebru2018datasheets,arnold2019factsheets}. However, knowing how to apply them effectively to the documentation needs of specific models for use by specific teams is challenging. It is difficult to know where to begin, what to include, and how to present complex model information in an accessible way \cite{richards2020methodology,holstein2019improving}.  We experienced this first-hand when tasked with helping different teams of AI model developers within our organization create AI documentation for a variety of model types. It was difficult for teams to generate the descriptive content required. Furthermore, it was hard to evaluate whether what they had written communicated in a manner that readers would consider to be "good."

Although there is a lack of research on how to write and evaluate documentation specifically for AI model developers, there is a large body of work that describes how to measure the quality of more general software documentation \cite{chomal2015software,sommerville2001software} along with various elicitation methods that can be used to extract knowledge from subject matter experts \cite{boose1990knowledge,agarwal1990knowledge}. This inspired us to explore the application of such methods to the task of creating AI documentation. 

In this paper, we propose and evaluate a set of quality dimensions to identify specific ways in which portions of AI documentation could be improved. Then, using those dimensions, we evaluate three different approaches for eliciting this content from model developers. We show how the dimensions identify shortcomings and propose how such dimensions can be used to further enable users to create useful AI documentation. Through these qualitative empirical studies, we make the following contributions: 

\begin{itemize}
    \item Contribution 1: An initial set straightforward and actionable set of quality dimensions suitable for multiple types of AI documentation.
    \item Contribution 2: A study that validates the usefulness of the dimensions including perspectives from 35 real-world practitioners who used them to evaluate AI documentation.
    \item Contribution 3: Real-world examples that shed light on how the quality dimensions support better documentation across multiple kinds of models and domains.
\end{itemize}

\section{Background and Related Work}

Despite considerable research on writing conventional software documentation, it remains hard to do those documents. Assessment of the quality of conventional documentation is also difficult, further complicating the task of improving the process of writing it. When AI, with its oftentimes opaque and probabilistic nature, is added into the mix, the difficulty understandably increases. In this work, we build on published approaches for measuring the quality of both software documentation and other forms of human communication to the area of documenting AI.

\subsection{Software Documentation}
Software documentation is a critical activity that goes beyond the development and creation of software and computational models \cite{kipyegen2013importance,parnas2011precise}. Documentation describes particular characteristics of a system, procedure, or software \cite{chomal2015software} and is shaped by the generally unique requirements of the project, product or model it is describing \cite{sommerville2001software}. Despite its necessity, the creation of documentation is disliked by programmers \cite{parnas2011precise}. There are several reasons for this. Developers and software engineers lack interest in the creation of documentation and its creation usually falls on technical writers who do not know all the details of the software they are documenting \cite{parnas2011precise}. Perhaps not surprisingly, end-users dislike the resulting documentation as it is often incomplete, difficult to read, and out of date \cite{chomal2015software}. Regardless, documentation remains a fundamental forms of communication between the developers of the software and the end-users \cite{kipyegen2013importance,chomal2015software}.  

Researchers have established the benefits and costs associated with the development, or lack thereof, of documentation \cite{sommerville2001software,kipyegen2013importance,zhi2015cost}. In addition to being a bridge between developers and end users, documentation can serve useful roles in problem determination, and evaluation of software capabilities prior to acquisition \cite{kipyegen2013importance}. The absence of reasonable documentation also makes the on-boarding of new team members more difficult and can lead developers down non-fruitful paths when debugging \cite{parnas2011precise}. 

\subsection{AI Documentation}

Today, the field of AI models is rapidly expanding into new application domains, leveraging new capabilities of ever more capable technology \cite{jordan2015machine}. Despite this growth, there is little research on the documentation that should accompany intelligent systems, or software that contains AI components, especially for users with different backgrounds who might be trying to consume the documentation \cite{gebru2018datasheets}. While several proposals have been made for \textit{what} should be included in this documentation, \cite{mitchell2019model,arnold2019factsheets}, little work has tried to assess the usefulness and quality of documentation that contains this information.

The present research focuses on the evaluation and assessment of AI model documentation. If successful, the underlying dimensions of quality and a standard approach to their assessment could provide ways to both improve documentation and compare its quality across models or over time. Earlier work focused on the strategies to create this type of documentation \cite{richards2020methodology,hind2020experiences} and the exploration of what type of content would be required to improve trust in AI models \cite{arnold2019factsheets}. 

The present study took a two part approach. The first part evaluated AI documentation quality using dimensions previously developed for assessing both conventional software documentation \cite{ding2014knowledge,zhi2015cost} and pragmatics \cite{mey2001pragmatics}. The second part examined whether the quality of AI documentation could be improved through either examples of good documentation or by asking writers to reflect on a series of questions that might induce them to write content that would be, for example, more complete, more concise, and more relevant. For this second part, we asked a group of experts with background in data science to evaluate the results of this writing using the quality measurements investigated in the first part of the study.

\subsection{Measures of Quality}

Software documentation quality research and communication research serve as the foundation for the quality dimensions that we explore in this paper.

\subsubsection{Software documentation quality}

Researchers have developed sets of rules \cite{parnas2011precise}, corollaries, and methodologies to classify and understand the qualities \cite{smart2002assessing,plosch2014value,ding2014knowledge} of software documentation.  Many of these sets of rules share similarities with each other. Quality attributes such as accuracy, concreteness, writing style, and understandability have been offered as useful dimensions of quality \cite{plosch2014value}. Other dimensions such as completeness, unambiguity, conciseness, and ease of access have been proposed \cite{parnas2011precise}. Yet other dimensions include consistency, traceability, reusability, format, trustworthiness, consistency, and retrievability \cite{ding2014knowledge,zhi2015cost}. All such proposals aim to be  useful in capturing what is essential to documentation quality \cite{kipyegen2013importance}.

\subsubsection{Communication quality}

Ultimately, software documentation's main goal is to communicate some idea to its reader. Thus, we also looked at the established literature in communication and its related field of pragmatics. Communication is inherently social and is best understood within the different contexts or purposes in which it occurs \cite{littlejohn2010theories}. A face to face conversation clearly differs from an internal memorandum but both have communicative intent. Software documentation also has a communication intent but this is often obscured. At heart though, documentation provides a channel to communicate important information about a product, system and/or model between the creators within a team over time or between the creators and users \cite{kipyegen2013importance}. As such, frameworks and theories that have been applied to an analysis of human communication might provide guidance in evaluating such documents. 

\section{Evaluation Dimensions}

Assessing and measuring the quality of products, models and software is not trivial. There are different definitions and standards of quality, each one of them with their own limitations and benefits \cite{smart2002assessing}. There have been several approaches to the definition of the dimensions of quality \cite{anderson1994cross,parasuraman1994alternative,garvin1987competing}. We wanted to define a series of quality dimensions that could evaluate the quality of the content in AI documentation. The dimensions needed to be grounded in prior work, easily understandable, actionable, and easy to use. We were particularly inspired by existing work on measuring quality of software documentation ~\cite{parnas2011precise,smart2002assessing,plosch2014value,ding2014knowledge,zhi2015cost} and measuring quality of conversations ~\cite{leech2016principles,mey2001pragmatics}. In software documentation studies we gave priority to quality dimensions that surfaced in more than one work. For example, completeness \cite{smart2002assessing,parnas2011precise}, accuracy \cite{smart2002assessing,parnas2011precise,plosch2014value}, and conciseness \cite{ding2014knowledge} are recurring quality dimensions for software documentation analyses and we decided to adapt them to the AI model documentation evaluation. Although, documentation does not typically hold a conversation with the user, we found the field of pragmatics to be particularly useful for capturing effective ways to communicate ideas. This application of pragmatics to AI is not without precedent as it has been used in the past to evaluate the quality of conversations with conversational agents ~\cite{sandbank2018detecting}.

Grice proposed four dimensions or maxims of conversational quality \cite{mey2001pragmatics}: 
\begin{itemize}
    \item \textit{Maxim of Quality:} Be truthful. Do not say what you believe to be false. Do not say that for which you lack adequate evidence.
    \item \textit{Maxim of Quantity:} Quantity of information. Make your contribution as informative as is required (for the current purposes of the exchange).
    \item \textit{Maxim of Relation:} Be relevant.
    \item \textit{Maxim of Manner:} Be clear. Avoid ambiguity.
\end{itemize}

We combined these maxims with the dimensions of quality arising from studies of software documentation to define the complete set of dimensions presented in Table  \ref{tbl:qualityMetrics}. Our goals were to create a set of dimensions that could point specific flaws in documentation and provide useful guidance for improving these flaws. The maxims also provided a useful higher order categorization of the full set of dimensions. Completeness, supporting evidence, vocabulary choice, and clarity fell within the maxim of quality. Conciseness fell within the maximum of quantity. Relevance and structure fell within the maxim of relation as a basis. Finally, the representation and navigability dimensions fell within the maxim of manner. As the maxims are known aspects of communication quality, good coverage of the four helped ensure that important quality dimensions of software documentation were not missed.

\footnotesize
\begin{table}[]
\caption{The quality dimensions for AI documentation}
\begin{tabular}{p{1in}p{3in}}
Dimension              & Definition                                                                                                                   \\ \toprule
Completeness        & This fact contains all the information necessary to understand its content. No information is missing.                       \\ \midrule
Conciseness         & This fact contains only information about its content. The fact conveys its content efficiently.                             \\ \midrule
Relevance           & This fact addresses the category/question it belongs to. All information is on-topic.                                        \\ \midrule
Supporting Evidence & If this fact requires evidence, it has sufficient evidence to support its claims.                                            \\ \midrule
Vocabulary Choice   & This fact's terminology and word choice is appropriate to its intended audience.                                             \\ \midrule
Clarity             & This fact's terminology and other content is sufficiently understandable. Nothing is ambiguous, obscure or incomprehensible. \\ \midrule
Structure           & This fact has a logical structure and information flow.                                                                      \\ \midrule
Representation      & This fact's presentation style (text, graph, table, etc.) is appropriate for its content.                                    \\ \midrule
Navigability        & This fact's representation makes it straightforward to find information of interest.                                        \\ \bottomrule
\label{tbl:qualityMetrics}
\end{tabular}
\end{table}
\normalsize
\section{Experimental Design}


To address our research questions, we completed two studies. The first study was an AI documentation \textit{evaluation} survey, completed by participants who had experience building AI models, assessing the quality of documentation that had been created previously. The goal of this study was to see if the dimensions were usable by data scientists and to see if the dimensions were considered to be useful or not by that audience. The second study was an AI documentation \textit{creation} exercise in which participants wrote a section of documentation for a model they recently worked on. The second study's goal was to see how different approaches for eliciting documentation affected documentation quality but also an exercise in determining if the dimensions could highlight specific, actionable issues in the resulting documentation.

\subsection{Study 1: Evaluating Documentation Quality Using the Dimensions}


\subsubsection{Participants}

We recruited participants who had experience building AI models from a large information technology company on internal Slack \footnote{https://slack.com} channels for data scientists. We required that participants had at least 1 year of data science experience. Survey participants were entered into a prize raffle worth approximately \$50. Our call for participants resulted in 35 model builders whose experience in data science ranged from 1 to 26 years with an average of 5.2 years.

\subsubsection{FactSheets as the AI Documentation}

As our source of AI documentation to evaluate, we used previously developed FactSheets ~\cite{arnold2019factsheets} freely available on the AI FactSheets 360 website\footnote{https://aifs360.mybluemix.net/examples}. At the time of the study, there were six FactSheets available, representing a diverse range of models and domains such as detecting breast cancer, generating captions for images, and classifying audio sounds.

The example FactSheets organized the content into separate \emph{facts}, which were simply sections of documentation on a particular topic such as a model's purpose or details about its performance. All the FactSheets had the same set of facts, enabling us to compare the same fact across multiple FactSheets. As the FactSheets were quite long, anywhere from 5 to 13 pages when printed, we selected only a subset of the facts for the purposes of the experiment. In particular, we were interested in including both documentation that had to be hand authored and documentation that could reasonably be automatically generated, as such documentation represents the two endpoints on the spectrum of how we have observed that such documentation is created. Our selection is presented in Table \ref{tbl:facts}.

\begin{figure*}[ht]
    \includegraphics[scale=0.2]{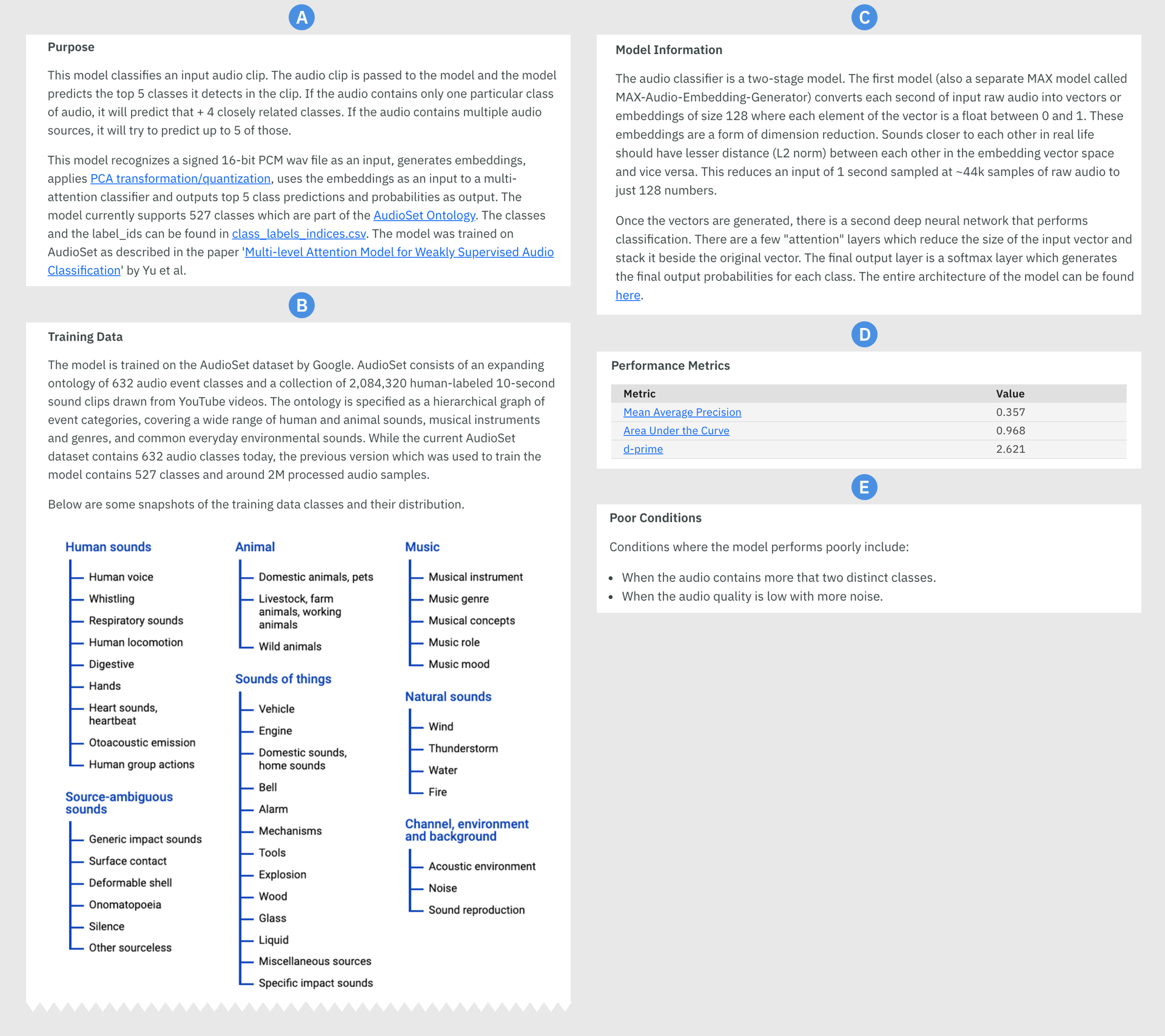}
    \centering
    \caption{Full AI documentation was shown in a long scrolling web document (not shown) that provided context for individual excerpts of AI documentation reviewed and written by study participants. Examples of excerpts from one document are shown in (A) Purpose, (B) Training Data, (C) Model Information, (D) Performance Metrics, and (E) Poor Conditions .}
    \label{fig:factsheet}
\end{figure*}

The facts that we selected are found in Table \ref{tbl:facts}. An example FactSheet is shown in Figure \ref{fig:factsheet}

\footnotesize
\begin{table}[]
\caption{The facts used for the evaluation survey}
\begin{tabular}{p{1in}p{3in}}
Fact                & Content                                                                                                      \\ \toprule
Purpose             & Information about the model's goal.                                                                           \\ \midrule
Model Information   & Information about the model's design and structure.                                                           \\ \midrule
Training Data       & Information about the training data (if appropriate).                                                         \\ \midrule
Performance Metrics & Measurements of the model's performance and quality.                                                          \\ \midrule
Poor Conditions     & Information about the circumstances where the model performs poorly (within expected use cases of the model).  \\ \bottomrule
\label{tbl:facts}
\end{tabular}
\end{table}
\normalsize

\subsubsection{Evaluation Details}

In the survey, participants were asked to read and rate three to five facts from one of the six FactSheets using the nine dimensions and their definitions from Table \ref{tbl:qualityMetrics}. Participants were asked to rate each fact for each dimension using a 4-point Likert Scale: "Very Poor" to "Very Good" with no neutral option. Additionally, participants were given the option to mark a dimension as "Not Applicable" if they did not believe the dimension was suitable for that fact.

In addition to this evaluation, we asked participants to select which dimensions they considered helpful in assessing the quality of the documentation and which ones were not helpful. Participants were also asked to detail the reasons why they were helpful or not and also given an opportunity to suggest additional quality dimensions.

\subsection{Study 2: Creating AI Documentation}

\subsubsection{Participants}

In Study 2, we recruited participants with experience building models from internal Slack channels. In addition to having at least one year of experience in data science, we also required that they had recently finished a model and were knowledgeable enough about it to create documentation for it. Our call resulted in 19 participants whose experience ranged from 2 to 10 years with an average of 5 years.

\subsubsection{AI documentation creation treatments}

Our goals for this study were to have participants create a portion of documentation for a wide variety of models and domains and then use the quality dimensions to assess the resulting documentation. We were also interested in whether we could encourage participants to write higher quality documentation, so we split the task into three treatments. The first treatment was the \emph{Baseline} condition where participants received no additional guidance for writing documentation. The second treatment was the \emph{Examples} treatment where we provided participants copies of the FactSheets used in the first study. The third and final treatment, the \emph{User-centric} condition, participants were asked to fill out some questions borrowed from the methodology to create Factsheets ~\cite{richards2020methodology}, that asked the user to explicitly think of the needs of the user they are writing the documentation for. We asked the following additional questions only in the User-centric condtion:
\footnotesize
\begin{enumerate}
    \item What are the most important pieces of information that you need to know for this section?
    \item What is <Persona's name ... see below> going to be asking for when looking at this section?
    \item Will there need to be additional definitions for terms that <Persona's name> is unfamiliar with in this section?
    \item What model-specific, domain-specific or other knowledge do you have that may not be obvious to others in this section?
\end{enumerate}
\normalsize

We ran the experiment remotely via WebEx\footnote{https://webex.com}. Participants shared their screen as they edited a document containing the task instructions for one of three treatments and spaces for their responses. Each participant was given 20 minutes to complete their documentation, but were allowed to finish early if they felt they were done. 

Participants were randomly assigned a section of documentation corresponding to one of the five facts that we used in Study 1 and asked to write their documentation as if in the context of a larger piece of documentation. The intent was to encourage participants to not include content that would reasonably be expected to appear in other sections. To help them reason about what may be included in the rest of the documentation, we provided a short description of each section's content as in Table \ref{tbl:facts}. The overall structure of the documentation contained the following sections: Model Name, Purpose, Intended Domain, Training Data, Model Information, Inputs and Outputs, Performance Metrics, Accuracy, Bias, Robustness, Domain Shift, Test Data, Optimal Conditions, Poor Conditions, and Explainability. Those who received the Examples or User-centric treatments were given an additional 10 minutes before creating the documentation to look over the examples and answer the questions.

\subsubsection{Persona details}

To help the user think about the person they were writing the documentation for, all three treatments included a persona named Carmen. The key features in this persona were that Carmen had six months of experience in data science, she understood basic machine learning models such as ``regressions, decision trees, SVMs, Naive Bayes and K-means'' but had ``no experience with deep learning or neural networks.'' Carmen was described as having a technical background but was fairly new to machine learning to encourage participants to write for someone less skilled than them, but still reasonably knowledgeable about data science. Carmen's learning style was described as ``a comprehensive information processing style, preferring to gather information and learn about a topic before jumping into it.'' This was intended to remind participants that Carmen may need additional explanations.

\section{Results}

\subsection{Study 1: Evaluating the Quality Dimensions}

\footnotesize
\begin{table}[]
\caption{Number of evaluations per fact per model for the quality evaluation}
\begin{tabular}{p{1.25in}p{0.55in}p{0.55in}p{0.55in}p{0.55in}p{0.55in}}
\thead{Model} &\thead{Model Information} & \thead{Performance Metrics} & \thead{Poor Conditions} & \thead{Purpose} & \thead{Training Data} \\
\toprule
Audio Classifier          & 4                 & 4                   & 4               & 4       & 4             \\
\midrule
Breast Cancer Detector    & 4                 & 4                   & 3               & 7       & 3             \\
\midrule
Image Caption Generator   & 1                 & 2                   & 2               & 1       & 3             \\
\midrule
Object Detector           & 5                 & 7                   & 6               & 10      & 6             \\
\midrule
Text Sentiment Classifier & 5                 & 3                   & 5               & 3       & 5             \\
\midrule
Weather Forecaster        & 3                 & 4                   & 4               & 4       & 5             \\
\bottomrule
\label{tbl:qualityScoreCounts}
\end{tabular}
\end{table}
\normalsize

Our 35 participants evaluated a total of 125 facts across the 6 models. A breakdown of how many facts were evaluated across all the FactSheets is shown in Table \ref{tbl:qualityScoreCounts}. 

\subsubsection{Quality dimensions evaluation}

\tiny

\begin{table}[]
\caption{Average quality scores grouped by model. Negative scores represent "Poor" scores and are highlighted}
\begin{tabular}{p{0.6in}p{0.5in}p{0.3in}p{0.3in}p{0.3in}p{0.3in}p{0.3in}p{0.3in}p{0.3in}p{0.3in}p{0.3in}}
Model                                       & Fact                & \rotatebox{60}{Completeness}                 & \rotatebox{60}{Conciseness}                 & \rotatebox{60}{Relevance}                   & \rotatebox{60}{Supporting Evidence}          & \rotatebox{60}{Vocabulary Choice}            & \rotatebox{60}{Clarity}                      & \rotatebox{60}{Structure}                   & \rotatebox{60}{Representation}               & \rotatebox{60}{Navigability}                 \\ \toprule
                                            & Model Info   & 1.5  & 1.3 & 1.5 & 1.5  & 1.3  & 1.3  & 1.0 & 0.5  & 1.3  \\
                                            & Perf. Metrics & 0.8  & 2.0 & 2.0 & 0.0  & 2.0  & 2.0  & 1.7 & 1.5  & 1.8  \\
                                            & Poor Conds.     & 0.3  & 1.3 & 1.0 & 0.8  & 1.3  & 0.7  & 1.5 & 0.7  & 1.5  \\
                                            & Purpose             & 1.3  & 1.8 & 1.5 & 2.0  & 1.5  & 1.5  & 1.8 & 1.8  & 1.5  \\
\multirow{-5}{0.6in}{Audio Classifier}          & Training Data       & 2.0  & 2.0 & 1.3 & 2.0  & 1.8  & 1.8  & 1.3 & 2.0  & 1.8  \\ \midrule
                                            & Model Info   & 1.8  & 1.3 & 1.8 & 1.3  & 1.5  & 1.5  & 1.3 & 0.8  & 0.8  \\
                                            & Perf. Metrics & 0.0  & 1.8 & 1.8 & 0.3  & 1.5  & 1.8  & 1.0 & 1.0  & 1.5  \\
                                            & Poor Conds.     & \cellcolor[HTML]{D2D2D4}-0.7 & 0.7 & 1.3 & \cellcolor[HTML]{BEBEBF}-1.0 & 0.7  & 0.7  & 0.7 & \cellcolor[HTML]{BEBEBF}-1.0 & 1.3  \\
                                            & Purpose             & 0.7  & 1.0 & 1.3 & 0.4  & 1.0  & 1.0  & 1.1 & 1.1  & 1.4  \\
\multirow{-5}{0.6in}{Breast Cancer Detection}   & Training Data       & 1.3  & 1.0 & 1.3 & 1.7  & 1.3  & 1.3  & 1.7 & 0.7  & 1.3  \\ \midrule
                                            & Model Info   & 0.6  & 1.4 & 0.8 & 0.8  & 1.2  & 1.4  & 1.0 & 1.4  & 0.8  \\
                                            & Perf. Metrics & 1.1  & 1.3 & 1.6 & 1.3  & 1.9  & 1.6  & 1.7 & 1.6  & 1.6  \\
                                            & Poor Conds.     & 0.0  & 1.3 & 0.5 & 0.7  & 1.5  & 1.0  & 1.5 & 1.3  & 1.0  \\
                                            & Purpose             & 1.3  & 1.4 & 1.4 & 0.7  & 1.5  & 1.3  & 1.3 & 1.2  & 1.2  \\
\multirow{-5}{0.6in}{Object Detector}           & Training Data       & 1.8  & 1.8 & 1.8 & 1.7  & 1.7  & 1.7  & 1.7 & 1.7  & 1.5  \\ \midrule
                                            & Model Info   & \cellcolor[HTML]{E3E3E5}-0.4 & 1.2 & 0.6 & 0.8  & \cellcolor[HTML]{EFEFF2}-0.2 & \cellcolor[HTML]{E3E3E5}-0.4 & 0.4 & 0.6  & 1.0  \\
                                            & Perf. Metrics & \cellcolor[HTML]{E7E7E9}-0.3 & 1.3 & 1.3 & 1.0  & 0.0  & 0.0  & 1.0 & 0.0  & 0.7  \\
                                            & Poor Conds.     & \cellcolor[HTML]{BEBEBF}-1.0 & 0.4 & 0.4 & \cellcolor[HTML]{A5A5A6}-1.4 & 0.4  & 0.0  & 0.4 & \cellcolor[HTML]{E3E3E5}-0.4 & \cellcolor[HTML]{EFEFF2}-0.2 \\
                                            & Purpose             & 1.3  & 1.0 & 0.7 & 0.3  & 1.3  & 1.0  & 0.3 & 0.3  & 1.0  \\
\multirow{-5}{0.6in}{Text Sentiment Classifier} & Training Data       & 0.4  & 1.6 & 1.4 & 0.0  & 0.4  & 0.0  & 1.6 & 0.8  & 1.0  \\ \midrule
                                            & Model Info   & 1.3  & 0.7 & 1.3 & \cellcolor[HTML]{E7E7E9}-0.3 & 0.7  & 1.3  & 1.3 & 1.0  & 1.3  \\
                                            & Perf. Metrics & 0.8  & 0.8 & 0.8 & 0.3  & 0.8  & 0.8  & 0.3 & 0.3  & 0.8  \\
                                            & Poor Conds.     & 0.3  & 0.8 & 0.5 & \cellcolor[HTML]{CDCDCF}-0.8 & 1.0  & 1.0  & 0.3 & 0.5  & 1.5  \\
                                            & Purpose             & 0.8  & 0.3 & 1.3 & \cellcolor[HTML]{ECECEF}-0.3 & 0.5  & 0.8  & 1.3 & 0.0  & 0.0  \\
\multirow{-5}{0.6in}{Weather Forecaster}        & Training Data       & 1.6  & 1.6 & 1.6 & 1.6  & 1.6  & 1.6  & 1.2 & 0.8  & 1.0 \\ \bottomrule
\label{tbl:qualityScoreByModel}
\end{tabular}
\end{table}

\begin{table}[]
\caption{Average quality scores grouped by model. High agreement scores in light green and high disagreement scores in medium red.}
\begin{tabular}{p{0.6in}p{0.5in}p{0.3in}p{0.3in}p{0.3in}p{0.3in}p{0.3in}p{0.3in}p{0.3in}p{0.3in}p{0.3in}}
Model & Fact & \rotatebox{60}{Completeness} & \rotatebox{60}{Conciseness} & \rotatebox{60}{Relevance} & \rotatebox{60}{Supporting Evidence} & \rotatebox{60}{Vocabulary Choice} & \rotatebox{60}{Clarity} & \rotatebox{60}{Structure} & \rotatebox{60}{Representation} & \rotatebox{60}{Navigability} \\ \toprule
 & Model Info & 1.50 & 1.25 & 1.50 & 1.50 & 1.25 & 1.25 & 1.00 & \cellcolor[HTML]{EBAB9B}{ 0.50} & 1.25 \\
 & Perf. Metrics & 0.75 & \cellcolor[HTML]{E2EFDA}{ 2.00} & \cellcolor[HTML]{E2EFDA}{ 2.00} & \cellcolor[HTML]{EBAB9B}{ 0.00} & \cellcolor[HTML]{E2EFDA}{ 2.00} & \cellcolor[HTML]{E2EFDA}{ 2.00} & 1.67 & 1.50 & 1.75 \\
 & Poor Conds. & 0.25 & 1.25 & 1.00 & 0.75 & 1.33 & \cellcolor[HTML]{EBAB9B}{ 0.67} & 1.50 & \cellcolor[HTML]{EBAB9B}{ 0.67} & 1.50 \\
 & Purpose & 1.25 & 1.75 & 1.50 & \cellcolor[HTML]{E2EFDA}{ 2.00} & 1.50 & 1.50 & 1.75 & 1.75 & 1.50 \\ 
\multirow{-5}{0.6in}{Audio Classifier} & Training Data & \cellcolor[HTML]{E2EFDA}{ 2.00} & \cellcolor[HTML]{E2EFDA}{ 2.00} & 1.67 & \cellcolor[HTML]{E2EFDA}{ 2.00} & 1.75 & 1.75 & 1.25 & \cellcolor[HTML]{E2EFDA}{ 2.00} & 1.75 \\ \midrule

 & Model Info & 1.75 & 1.25 & 1.75 & 1.67 & 1.50 & 1.50 & 1.25 & 0.75 & 0.75 \\
 & Perf. Metrics & \cellcolor[HTML]{EBAB9B}{ 0.00} & 1.75 & 1.75 & \cellcolor[HTML]{EBAB9B}{ 0.33} & 1.50 & 1.75 & 1.33 & 1.00 & 1.50 \\
 & Poor Conds. & \cellcolor[HTML]{EBAB9B}{ -0.67} & \cellcolor[HTML]{EBAB9B}{ 0.67} & 1.33 & \cellcolor[HTML]{EBAB9B}{-1.00} & \cellcolor[HTML]{EBAB9B}{ 0.67} & \cellcolor[HTML]{EBAB9B}{ 0.67} & 1.00 & \cellcolor[HTML]{EBAB9B}{-1.00} & 1.33 \\
 & Purpose & 0.71 & 1.00 & \cellcolor[HTML]{E2EFDA}{ 1.29} & 0.60 & 1.00 & 1.00 & \cellcolor[HTML]{E2EFDA}{ 1.14} & 1.33 & 1.43 \\ 
 \multirow{-5}{0.6in}{Breast Cancer Detection} & Training Data & 1.33 & 1.00 & 1.33 & 1.67 & 1.33 & 1.33 & 1.67 & \cellcolor[HTML]{EBAB9B}{ 0.67} & 1.33 \\ \midrule

 & Model Info & 0.75 & 1.75 & 1.00 & 1.00 & 1.50 & 1.75 & 1.25 & 1.75 & 1.00 \\
 & Perf. Metrics & 1.14 & 1.29 & 1.57 & 1.29 & \cellcolor[HTML]{E2EFDA}{ 1.86} & 1.57 & \cellcolor[HTML]{E2EFDA}{ 1.71} & \cellcolor[HTML]{E2EFDA}{ 1.83} & \cellcolor[HTML]{E2EFDA}{ 1.83} \\
 & Poor Conds. & 0.00 & 1.33 & \cellcolor[HTML]{EBAB9B}{ 0.50} & 0.80 & 1.50 & 1.00 & 1.50 & 1.60 & 1.20 \\
 & Purpose & 1.30 & 1.40 & 1.40 & 1.17 & 1.67 & 1.44 & 1.44 & \cellcolor[HTML]{E2EFDA}{ 1.71} & 1.33 \\ 
\multirow{-5}{0.6in}{Object Detector} & Training Data & \cellcolor[HTML]{E2EFDA}{ 1.83} & \cellcolor[HTML]{E2EFDA}{ 1.83} & \cellcolor[HTML]{E2EFDA}{ 1.83} & \cellcolor[HTML]{E2EFDA}{ 2.00} & \cellcolor[HTML]{E2EFDA}{ 2.00} & \cellcolor[HTML]{E2EFDA}{ 2.00} & \cellcolor[HTML]{E2EFDA}{ 2.00} & \cellcolor[HTML]{E2EFDA}{ 2.00} & \cellcolor[HTML]{E2EFDA}{ 1.80} \\ \midrule

 & Model Info & -0.40 & \cellcolor[HTML]{E2EFDA}{ 1.20} & \cellcolor[HTML]{EBAB9B}{ 0.60} & \cellcolor[HTML]{EBAB9B}{1.00} & \cellcolor[HTML]{EBAB9B}{ -0.20} & \cellcolor[HTML]{EBAB9B}{ -0.40} & 0.40 & \cellcolor[HTML]{EBAB9B}{ 0.60} & 1.00 \\
 & Perf. Metrics & -0.33 & 1.33 & 1.33 & 1.00 & \cellcolor[HTML]{EBAB9B}{ 0.00} & \cellcolor[HTML]{EBAB9B}{ 0.00} & 1.00 & \cellcolor[HTML]{EBAB9B}{ 0.00} & 1.00 \\
 & Poor Conds. & -1.00 & 0.40 & 0.40 & -1.40 & \cellcolor[HTML]{EBAB9B}{ 0.40} & 0.00 & 0.40 & -0.40 & \cellcolor[HTML]{EBAB9B}{ -0.20} \\
 & Purpose & 1.33 & 1.00 & \cellcolor[HTML]{EBAB9B}{ 0.67} & \cellcolor[HTML]{EBAB9B}{1.00} & 1.33 & 1.00 & 0.33 & 0.33 & 1.50 \\ 
\multirow{-5}{0.6in}{Text Sentiment Classifier} & Training Data & 0.40 & 1.60 & 1.40 & \cellcolor[HTML]{EBAB9B}{ 0.00} & 0.40 & 0.00 & 1.60 & 0.80 & 1.00 \\ \midrule

 & Model Info & 1.33 & \cellcolor[HTML]{EBAB9B}{ 0.67} & 1.33 & \cellcolor[HTML]{EBAB9B}{ -0.33} & \cellcolor[HTML]{EBAB9B}{ 0.67} & 1.33 & 1.33 & 1.00 & 1.33 \\
 & Perf. Metrics & 0.75 & 0.75 & 0.75 & 0.25 & 0.75 & 0.75 & 0.25 & 0.25 & 0.75 \\
 & Poor Conds. & 0.25 & 0.75 & 0.50 & -1.00 & 1.00 & 1.00 & 0.33 & \cellcolor[HTML]{EBAB9B}{ 0.67} & 1.50 \\
 & Purpose & 0.75 & 0.25 & 1.25 & -0.33 & 0.50 & 0.75 & 1.25 & 0.00 & 0.00 \\
\multirow{-5}{0.6in}{Weather Forecaster} & Training Data & 1.60 & 1.60 & 1.60 & 1.60 & 1.60 & 1.60 & 1.20 & 0.80 & 1.00 \\ \bottomrule
\label{tbl:agreement}
\end{tabular}
\end{table}

\normalsize

An initial glance at average scores across scorers as shown in Table \ref{tbl:qualityScoreByModel}, with scores in the "Poor" range highlighted, seem to indicate that the Audio Classifier and Object Detector models are of a higher quality than the other models. (We omit the Image Caption Generator from this table due to it receiving fewer scores than the other models.) However, this approach omits the views and scores of the individual scorers who may disagree with the majority of the scores. Instead, we need to look at a combination of both the score and its variance to address this problem. 

Table \ref{tbl:agreement} shows the average score across all scorers for each FactSheet and set of Facts, but also highlights the cells that have scores with very high levels and very low levels of consensus. This consensus was defined in terms of the score and variance for each of the facts.  In this table, cells highlighted in lights green are scores with a standard deviation $\sigma < 0.5$ indicating responses with a high level of agreement and a score in the range $[-2,-1)$ or $(1,2]$. Cells highlighted in a medium red have $\sigma > 1.5$ and an average score in the range $[-1,1]$, indicating very low levels of agreement.

With this more nuanced view, we can now see which Facts in the FactSheet were indeed "good" for a variety of participants with different skills and backgrounds. One standout is the Training Data fact for the Object detector model, which scored highly in every quality dimension. Each of the 6 scorers agreed that the Training Data fact for this model met their quality expectations. More importantly though, we can see the ones that split opinions among the scorers. Both the Model Information for the Text Sentiment Classifier and the Poor Conditions for the Breast Cancer Detection have large standard deviations. Similarly, what looked like a "good" model at first, the Audio Classifier Model too has some facts highlighted in red that deserve a closer look. These contested quality metrics are good candidates for a follow up with those scorers who gave it a low score to better understand the ways that those facts did not meet expectations.

\subsubsection{Participants opinions on the usefulness of the quality dimensions}

\footnotesize
\begin{table}[]
\caption{The number of participants who marked quality dimensions as useful or not}
\begin{tabular}{p{1.25in}p{.55in}p{.55in}}
Quality Dimension      & Useful Votes & Not Useful Votes \\ \toprule
Completeness        & 28           & 2                \\ \midrule
Supporting Evidence & 27           & 5                \\ \midrule
Clarity             & 26           & 1                \\ \midrule
Relevance           & 26           & 2                \\ \midrule
Conciseness         & 26           & 4                \\ \midrule
Representation      & 24           & 5                \\ \midrule
Navigability        & 21           & 16              \\ \midrule
Vocabulary Choice   & 20           & 7                \\ \midrule
Structure           & 19           & 8                \\ \bottomrule
\label{tbl:usefulMetrics}
\end{tabular}
\end{table}
\normalsize

On the whole, a majority of the 35 participants considered each of the nine quality dimensions useful, as seen in Table \ref{tbl:usefulMetrics}. Of these, the top-five most useful were Completeness (28 votes), Supporting Evidence (27 votes), Clarity (26 votes), Relevance (26 votes), and Conciseness (26 votes). On the not useful side, nearly half of participants considered Navigability to not be useful (16 votes). Although participants were given an opportunity to provide feedback as to why they chose dimensions as useful or not, only one participant clarified why they thought navigability was not helpful. That participant's comment was that navigability was more appropriate for the complete documentation instead of a single a small section such as a fact.

\subsubsection{Potential gaps and additional dimensions to consider}

In addition to scoring the quality dimensions we provided, we also asked participants to come up with any additional dimensions that they felt were missing. 14 of 35 participants suggested additional ones. They are aggregated and paraphrased in Table \ref{tbl:additionalQualityMetrics}.

Of the ones that participants suggested, several are closely related to the ones that we curated. For example, the suggestions of Comprehensibility (Is this information understandable to its intended audience?) and Definitions (Does the information define unfamiliar domain terminology and context?) are closely tied to the knowledge and skills of the documentation consumer, which is precisely the intent behind the dimension of Vocabulary Choice. However, two of the suggestions surprised us and did not immediately seem to relate to the dimensions we curated.

The dimension of Faithfulness, is ultimately a questions of trust. How can a reader of this documentation be sure that it is trustworthy? Outside of some authority that guarantees the veracity of the data within, this dimension is perhaps most closely related to AI governance. AI governance is a process in which AI model and application building lifecycles are closely monitored and surrounded by enforced rules to ensure specific business or regulatory guidelines are met. Trustworthy, or faithful documentation, would emerge, in part, from such a process.

Similarity to other documentation is an interesting way to consider the quality of a piece of documentation. This proposed dimension may be indicative of the uncertainty that seems to exist with practitioners that we spoke with about not being sure what to include in documentation for AI models. Certainly, being able to see multiple examples of documentation helps situate how "good" a given document is, and perhaps more importantly, highlights what it may be missing or what portions are particularly in need of improvement.

\subsubsection{Limitations}
The following limitations in the study could have affected the results obtained in the survey. Participants were not obligated to read the AI documentation in a specific time frame and were not constrained to finish the survey in a delimited amount of time. Consequently, the time they spent analyzing and reviewing the quality of the sections was not determined. Although the survey had a time set to know when they began, the participants could leave the survey open and could do the survey from start to end in a particular moment or solve the questions whenever they wanted to. Other important limitations of the study were the number of sections they had to analyze. Some participants could have become tired and evaluate with less attention the later sections in the survey.

\footnotesize
\begin{table}[]
\caption{Participants' suggestions for additional quality dimensions}
\begin{tabular}{p{1in}p{2.5in}p{1.5in}}
Suggested Dimension  & Definition                                                                    & Related dimensions                    \\ \toprule
Correctness       & How accurate is the information presented?                                    & Supporting Evidence                \\ \midrule
Balance           & Is there a suitable mix of information available? Will it fit in my head?     & Completeness, Conciseness, Clarity \\  \midrule
Comprehensibility & Is this information understandable to its intended audience?                  & Vocabulary Choice, Clarity         \\ \midrule
Faithfulness      & Is this information trustworthy?                                              &                                    \\ \midrule
Similarity        & How does this information compared to other documentation for similar models? &                                    \\ \midrule
Context           & Is this information contextually relevant?                                    & Relevance                          \\ \midrule
Length            & Is this information have the right amount of information?                     & Completeness, Conciseness          \\ \midrule
Definitions       & Does the information define unfamiliar domain terminology and context?        & Vocabulary Choice, Clarity         \\ \midrule
Success     & Does the information answer the needs of the reader?                           & Completeness, Relevance           \\ \bottomrule
\label{tbl:additionalQualityMetrics}
\end{tabular}
\end{table}
\normalsize

\subsection{Study 2: AI Documentation Creation}

Our 19 participants had developed models of several types for multiple domains. This diversity gives us some confidence in the generality of our observations. Model types included linear regressions, decision trees, convolutional neural networks and hidden Markov models. Problem domains included media, information technology, natural language processing, games, sales, and medicine.

\subsubsection{Evaluation of documentation}

Recall that our goal was to evaluate the documentation provided by participants using the quality dimensions and also to see if the support provided by Examples or the User-centric treatments would improve documentation quality. Our random split ended up with 7 participants in the Unassisted condition, 6 participants in the Examples condition and 6 participants in the User-centric condition. 

To evaluate the quality of the documentation we again used the quality dimensions to evaluate the documentation. This time, we asked experts in AI documentation to be our evaluators. All three evaluators (2 colleagues, 1 co-author) had written multiple examples of AI documentation equivalent to or more comprehensive than the FactSheets we used in the first study. Before the evaluation, each evaluator was given a 30-minute explanation of the quality dimensions, the context of the larger document within which the created facts would be contained, and the persona that the participants considered in writing their fact. The evaluators were asked to keep their quality dimension scores with the persona and larger document context in mind.

The results of the evaluation, represented as a majority vote between the three evaluators are presented in Table \ref{tbl:evalResults}. We also evaluated the scores using averages between the three evaluators, but they showed essentially identical results. Pairwise inter-rater reliabilities for the pairs were 69\%, 70\% and 76\%. Although there are some indicators that the User-centric condition may have resulted in overall lower quality documentation than Unassisted or Example, when we ran Kruskal-Wallis tests using the evaluators scores for each of the quality dimensions, we found no significant difference between the three treatments.

\footnotesize
\begin{table}[]
\caption{Documentation Creation Evaluation Results (Majority Vote)}
\begin{tabular}{r ccccccc|cccccc|cccccc}
Treatment           & \multicolumn{7}{c}{Unassisted}                                                                                                                                                                                                                                                                                                                                     & \multicolumn{6}{c}{Examples}                                                                                                                                                                                                                                                                                    & \multicolumn{6}{c}{User-centric}                                                                                                                                                                                                                                                                                \\ \toprule
Participant         & \multicolumn{1}{l}{5}                            & \multicolumn{1}{l}{7}                            & \multicolumn{1}{l}{9}                            & \multicolumn{1}{l}{10}                           & \multicolumn{1}{l}{13}                           & \multicolumn{1}{l}{16}                           & \multicolumn{1}{l}{19}                           & \multicolumn{1}{l}{2}                            & \multicolumn{1}{l}{4}                            & \multicolumn{1}{l}{6}                            & \multicolumn{1}{l}{8}                            & \multicolumn{1}{l}{14}                           & \multicolumn{1}{l}{18}                           & \multicolumn{1}{l}{1}                            & \multicolumn{1}{l}{3}                            & \multicolumn{1}{l}{11}                           & \multicolumn{1}{l}{12}                           & \multicolumn{1}{l}{15}                           & \multicolumn{1}{l}{17}                           \\ \midrule
Completeness        & \cellcolor[HTML]{C6EFCE}{\color[HTML]{006100} +} & \cellcolor[HTML]{FFC7CE}{\color[HTML]{9C0006} -} & \cellcolor[HTML]{C6EFCE}{\color[HTML]{006100} +} & \cellcolor[HTML]{C6EFCE}{\color[HTML]{006100} +} & \cellcolor[HTML]{C6EFCE}{\color[HTML]{006100} +} & \cellcolor[HTML]{C6EFCE}{\color[HTML]{006100} +} & \cellcolor[HTML]{FFC7CE}{\color[HTML]{9C0006} -} & \cellcolor[HTML]{C6EFCE}{\color[HTML]{006100} +} & \cellcolor[HTML]{C6EFCE}{\color[HTML]{006100} +} & \cellcolor[HTML]{FFC7CE}{\color[HTML]{9C0006} -} & \cellcolor[HTML]{C6EFCE}{\color[HTML]{006100} +} & \cellcolor[HTML]{C6EFCE}{\color[HTML]{006100} +} & \cellcolor[HTML]{FFC7CE}{\color[HTML]{9C0006} -} & \cellcolor[HTML]{FFC7CE}{\color[HTML]{9C0006} -} & \cellcolor[HTML]{C6EFCE}{\color[HTML]{006100} +} & \cellcolor[HTML]{FFC7CE}{\color[HTML]{9C0006} -} & \cellcolor[HTML]{C6EFCE}{\color[HTML]{006100} +} & \cellcolor[HTML]{FFC7CE}{\color[HTML]{9C0006} -} & \cellcolor[HTML]{FFC7CE}{\color[HTML]{9C0006} -} \\
Conciseness         & \cellcolor[HTML]{C6EFCE}{\color[HTML]{006100} +} & \cellcolor[HTML]{C6EFCE}{\color[HTML]{006100} +} & \cellcolor[HTML]{C6EFCE}{\color[HTML]{006100} +} & \cellcolor[HTML]{C6EFCE}{\color[HTML]{006100} }  & \cellcolor[HTML]{C6EFCE}{\color[HTML]{006100} +} & \cellcolor[HTML]{C6EFCE}{\color[HTML]{006100} +} & \cellcolor[HTML]{FFC7CE}{\color[HTML]{9C0006} -} & \cellcolor[HTML]{C6EFCE}{\color[HTML]{006100} +} & \cellcolor[HTML]{FFC7CE}{\color[HTML]{9C0006} -} & \cellcolor[HTML]{C6EFCE}{\color[HTML]{006100} +} & \cellcolor[HTML]{C6EFCE}{\color[HTML]{006100} +} & \cellcolor[HTML]{C6EFCE}{\color[HTML]{006100} +} & \cellcolor[HTML]{C6EFCE}{\color[HTML]{006100} +} & \cellcolor[HTML]{C6EFCE}{\color[HTML]{006100} +} & \cellcolor[HTML]{C6EFCE}{\color[HTML]{006100} +} & \cellcolor[HTML]{FFC7CE}{\color[HTML]{9C0006} -} & \cellcolor[HTML]{C6EFCE}{\color[HTML]{006100} +} & \cellcolor[HTML]{FFC7CE}{\color[HTML]{9C0006} -} & \cellcolor[HTML]{FFC7CE}{\color[HTML]{9C0006} -} \\
Relevance           & \cellcolor[HTML]{C6EFCE}{\color[HTML]{006100} +} & \cellcolor[HTML]{FFC7CE}{\color[HTML]{9C0006} -} & \cellcolor[HTML]{C6EFCE}{\color[HTML]{006100} +} & \cellcolor[HTML]{C6EFCE}{\color[HTML]{006100} +} & \cellcolor[HTML]{C6EFCE}{\color[HTML]{006100} +} & \cellcolor[HTML]{C6EFCE}{\color[HTML]{006100} +} & \cellcolor[HTML]{FFC7CE}{\color[HTML]{9C0006} -} & \cellcolor[HTML]{C6EFCE}{\color[HTML]{006100} +} & \cellcolor[HTML]{FFC7CE}{\color[HTML]{9C0006} -} & \cellcolor[HTML]{C6EFCE}{\color[HTML]{006100} +} & \cellcolor[HTML]{C6EFCE}{\color[HTML]{006100} +} & \cellcolor[HTML]{C6EFCE}{\color[HTML]{006100} +} & \cellcolor[HTML]{C6EFCE}{\color[HTML]{006100} +} & \cellcolor[HTML]{C6EFCE}{\color[HTML]{006100} +} & \cellcolor[HTML]{C6EFCE}{\color[HTML]{006100} +} & \cellcolor[HTML]{FFC7CE}{\color[HTML]{9C0006} -} & \cellcolor[HTML]{C6EFCE}{\color[HTML]{006100} +} & \cellcolor[HTML]{FFC7CE}{\color[HTML]{9C0006} -} & \cellcolor[HTML]{FFC7CE}{\color[HTML]{9C0006} -} \\
Supporting Evidence & \cellcolor[HTML]{C6EFCE}{\color[HTML]{006100} +} & 0                                                & \cellcolor[HTML]{FFC7CE}{\color[HTML]{9C0006} -} & \cellcolor[HTML]{C6EFCE}{\color[HTML]{006100} +} & \cellcolor[HTML]{C6EFCE}{\color[HTML]{006100} +} & 0                                                & \cellcolor[HTML]{FFC7CE}{\color[HTML]{9C0006} -} & \cellcolor[HTML]{C6EFCE}{\color[HTML]{006100} +} & \cellcolor[HTML]{C6EFCE}{\color[HTML]{006100} +} & \cellcolor[HTML]{C6EFCE}{\color[HTML]{006100} +} & \cellcolor[HTML]{C6EFCE}{\color[HTML]{006100} +} & \cellcolor[HTML]{FFC7CE}{\color[HTML]{9C0006} -} & \cellcolor[HTML]{FFC7CE}{\color[HTML]{9C0006} -} & \cellcolor[HTML]{FFC7CE}{\color[HTML]{9C0006} -} & \cellcolor[HTML]{FFC7CE}{\color[HTML]{9C0006} -} & \cellcolor[HTML]{FFC7CE}{\color[HTML]{9C0006} -} & \cellcolor[HTML]{C6EFCE}{\color[HTML]{006100} +} & \cellcolor[HTML]{FFC7CE}{\color[HTML]{9C0006} -} & \cellcolor[HTML]{FFC7CE}{\color[HTML]{9C0006} -} \\
Vocabulary Choice   & \cellcolor[HTML]{C6EFCE}{\color[HTML]{006100} +} & \cellcolor[HTML]{FFC7CE}{\color[HTML]{9C0006} -} & \cellcolor[HTML]{C6EFCE}{\color[HTML]{006100} +} & \cellcolor[HTML]{C6EFCE}{\color[HTML]{006100} +} & \cellcolor[HTML]{C6EFCE}{\color[HTML]{006100} +} & \cellcolor[HTML]{C6EFCE}{\color[HTML]{006100} +} & \cellcolor[HTML]{FFC7CE}{\color[HTML]{9C0006} -} & \cellcolor[HTML]{C6EFCE}{\color[HTML]{006100} +} & \cellcolor[HTML]{C6EFCE}{\color[HTML]{006100} +} & \cellcolor[HTML]{C6EFCE}{\color[HTML]{006100} +} & \cellcolor[HTML]{C6EFCE}{\color[HTML]{006100} +} & \cellcolor[HTML]{C6EFCE}{\color[HTML]{006100} +} & \cellcolor[HTML]{C6EFCE}{\color[HTML]{006100} +} & \cellcolor[HTML]{FFC7CE}{\color[HTML]{9C0006} -} & \cellcolor[HTML]{C6EFCE}{\color[HTML]{006100} +} & \cellcolor[HTML]{C6EFCE}{\color[HTML]{006100} +} & \cellcolor[HTML]{C6EFCE}{\color[HTML]{006100} +} & \cellcolor[HTML]{FFC7CE}{\color[HTML]{9C0006} -} & \cellcolor[HTML]{FFC7CE}{\color[HTML]{9C0006} -} \\
Clarity             & \cellcolor[HTML]{C6EFCE}{\color[HTML]{006100} +} & \cellcolor[HTML]{FFC7CE}{\color[HTML]{9C0006} -} & \cellcolor[HTML]{C6EFCE}{\color[HTML]{006100} +} & \cellcolor[HTML]{C6EFCE}{\color[HTML]{006100} +} & \cellcolor[HTML]{C6EFCE}{\color[HTML]{006100} +} & \cellcolor[HTML]{C6EFCE}{\color[HTML]{006100} +} & \cellcolor[HTML]{FFC7CE}{\color[HTML]{9C0006} -} & \cellcolor[HTML]{C6EFCE}{\color[HTML]{006100} +} & \cellcolor[HTML]{C6EFCE}{\color[HTML]{006100} +} & \cellcolor[HTML]{FFC7CE}{\color[HTML]{9C0006} -} & \cellcolor[HTML]{C6EFCE}{\color[HTML]{006100} +} & \cellcolor[HTML]{C6EFCE}{\color[HTML]{006100} +} & \cellcolor[HTML]{C6EFCE}{\color[HTML]{006100} +} & \cellcolor[HTML]{FFC7CE}{\color[HTML]{9C0006} -} & \cellcolor[HTML]{C6EFCE}{\color[HTML]{006100} +} & \cellcolor[HTML]{C6EFCE}{\color[HTML]{006100} +} & \cellcolor[HTML]{C6EFCE}{\color[HTML]{006100} +} & \cellcolor[HTML]{FFC7CE}{\color[HTML]{9C0006} -} & \cellcolor[HTML]{FFC7CE}{\color[HTML]{9C0006} -} \\
Structure           & \cellcolor[HTML]{C6EFCE}{\color[HTML]{006100} +} & 0                                                & \cellcolor[HTML]{C6EFCE}{\color[HTML]{006100} +} & \cellcolor[HTML]{C6EFCE}{\color[HTML]{006100} +} & \cellcolor[HTML]{C6EFCE}{\color[HTML]{006100} +} & \cellcolor[HTML]{C6EFCE}{\color[HTML]{006100} +} & \cellcolor[HTML]{FFC7CE}{\color[HTML]{9C0006} -} & \cellcolor[HTML]{C6EFCE}{\color[HTML]{006100} +} & \cellcolor[HTML]{FFC7CE}{\color[HTML]{9C0006} -} & \cellcolor[HTML]{FFC7CE}{\color[HTML]{9C0006} -} & \cellcolor[HTML]{C6EFCE}{\color[HTML]{006100} +} & \cellcolor[HTML]{C6EFCE}{\color[HTML]{006100} +} & \cellcolor[HTML]{C6EFCE}{\color[HTML]{006100} +} & \cellcolor[HTML]{FFC7CE}{\color[HTML]{9C0006} -} & \cellcolor[HTML]{C6EFCE}{\color[HTML]{006100} +} & \cellcolor[HTML]{FFC7CE}{\color[HTML]{9C0006} -} & \cellcolor[HTML]{C6EFCE}{\color[HTML]{006100} +} & \cellcolor[HTML]{FFC7CE}{\color[HTML]{9C0006} -} & \cellcolor[HTML]{FFC7CE}{\color[HTML]{9C0006} -} \\
Representation      & \cellcolor[HTML]{C6EFCE}{\color[HTML]{006100} +} & \cellcolor[HTML]{C6EFCE}{\color[HTML]{006100} +} & \cellcolor[HTML]{FFC7CE}{\color[HTML]{9C0006} -} & \cellcolor[HTML]{C6EFCE}{\color[HTML]{006100} +} & \cellcolor[HTML]{C6EFCE}{\color[HTML]{006100} +} & \cellcolor[HTML]{C6EFCE}{\color[HTML]{006100} +} & \cellcolor[HTML]{FFC7CE}{\color[HTML]{9C0006} -} & \cellcolor[HTML]{C6EFCE}{\color[HTML]{006100} +} & \cellcolor[HTML]{C6EFCE}{\color[HTML]{006100} +} & \cellcolor[HTML]{C6EFCE}{\color[HTML]{006100} +} & \cellcolor[HTML]{C6EFCE}{\color[HTML]{006100} +} & \cellcolor[HTML]{C6EFCE}{\color[HTML]{006100} +} & 0                                                & \cellcolor[HTML]{FFC7CE}{\color[HTML]{9C0006} -} & \cellcolor[HTML]{FFC7CE}{\color[HTML]{9C0006} -} & \cellcolor[HTML]{C6EFCE}{\color[HTML]{006100} +} & \cellcolor[HTML]{C6EFCE}{\color[HTML]{006100} +} & \cellcolor[HTML]{FFC7CE}{\color[HTML]{9C0006} -} & \cellcolor[HTML]{FFC7CE}{\color[HTML]{9C0006} -} \\
Navigability        & \cellcolor[HTML]{C6EFCE}{\color[HTML]{006100} +} & 0                                                & \cellcolor[HTML]{C6EFCE}{\color[HTML]{006100} +} & \cellcolor[HTML]{C6EFCE}{\color[HTML]{006100} +} & \cellcolor[HTML]{C6EFCE}{\color[HTML]{006100} +} & \cellcolor[HTML]{C6EFCE}{\color[HTML]{006100} +} & \cellcolor[HTML]{FFC7CE}{\color[HTML]{9C0006} -} & \cellcolor[HTML]{C6EFCE}{\color[HTML]{006100} +} & 0                                                & \cellcolor[HTML]{C6EFCE}{\color[HTML]{006100} +} & \cellcolor[HTML]{C6EFCE}{\color[HTML]{006100} +} & \cellcolor[HTML]{C6EFCE}{\color[HTML]{006100} +} & \cellcolor[HTML]{C6EFCE}{\color[HTML]{006100} +} & \cellcolor[HTML]{FFC7CE}{\color[HTML]{9C0006} -} & \cellcolor[HTML]{C6EFCE}{\color[HTML]{006100} +} & 0                                                & \cellcolor[HTML]{C6EFCE}{\color[HTML]{006100} +} & \cellcolor[HTML]{FFC7CE}{\color[HTML]{9C0006} -} & \cellcolor[HTML]{FFC7CE}{\color[HTML]{9C0006} -} \\ 
\bottomrule
\label{tbl:evalResults}
\end{tabular}
\end{table}
\normalsize

\subsubsection{Examples of how the quality dimensions reveal quality issues}

One additional measure of the usefulness of the quality dimensions is how well they help to identify issues. Here we present some examples of documentation from the second study and commentary provided by the evaluators.

\paragraph{Example 1}
\begin{quote}
    P4 (Purpose): ``The model aims to predict the handwritten digit or a character from a picture. The model is a CNN based model with the ability to detect 62 different classes (26 A-Z, 26 a-z, and 10 digits). The model is trained through supervised learning in a federated learning setting [insert reference], which does not require collecting the individual users data and instead keeps it securely on their own machines. A total of ~3000 users/clients participated in this training regime with an average number of data points on each client equal to 150 points. The model currently being served was trained for ~1000 communication rounds and reached ~90\% mean accuracy over all participating clients.''
\end{quote}

Evaluator 2 responded, ``Starts off well but segues into details that are not relevant to this section.''

\vspace{.25cm}
This was not uncommon. Another observed pattern was to provide additional context at the beginning which would have naturally been covered in another section of the document.

\paragraph{Example 2}
\begin{quote}
    P7 (Purpose): ``GNN based model can help us get the relationship with each node and their neighbours which can help us generate code summary better.''
\end{quote} 

Evaluator 2 responded, ``Sounds too abstract and [especially] from the perspective of the persona, incomprehensible.''

\vspace{.25cm}
While not too many examples are as terse as this, several others omitted key details and would leave the consumer, in this case Carmen, confused.

\paragraph{Example 3}
\begin{quote}
    P15 (Performance Metrics) ``On-policy first-visit Monte Carlo control using a soft epsilon policy can be used to play BlackJack. The method determines the optimal policy using the epsilon greedy method. The main input is the and open AI gym objects that simulate a blackjack dealer. The policy obtained is the ones that win the maximum number of games.''
\end{quote}

Evaluator 2 said, ``It uses a lot of Reinforcement Learning (RL) specific terminology. While I understand RL well, the targeted persona would have a very hard time understanding it.''

\vspace{.25cm}
This is a particularly difficult problem for many of our participants to avoid; The desire to use terminology that has specific technical meanings is very strong, and generally appropriate. But doing so should cause the writer to provide either a link to explanatory material or additional information within the fact itself.

\subsubsection{Limitations}
There were some important limitations in the present study that could have affected the results obtained. As participants wrote their fact in isolation from the rest of the documentation, they may have been drawn to provide context that would have normally appeared elsewhere in the FactSheet. If so, conciseness and relevance metrics would be particularly depressed.

\section{Conclusions}

Two high-level results stand out. The first is the overall positive assessment of the metrics as being useful. This is encouraging as the dimensions appeared useful to people with quite varied backgrounds in data science and software development. Among the dimensions, only navigability looked, to slightly less than half the participants, as \textit{not} useful. This is likely due to the fact that navigability is a more interesting property of an entire FactSheet as opposed to a single fact.
    
The second result is that the dimensions provided a good degree of differentiation between each other within a fact, and between facts, pointing to the likely utility of applying these dimensions to find areas of weakness in AI documentation that could be improved. In short, the dimensions appear to point to potential actions.
    
Our results highlight an ongoing need to find ways to elicit high quality facts. No elicitation method resulted in all associated facts being evaluated by our experts as "good". Providing high-quality examples appears to be somewhat beneficial. But asking people to reflect on questions such as supporting evidence, clarity, structure and so on did not measurably increase fact quality, and may even have lead to lower quality than the unassisted condition. 

Measuring the quality of documentation is critical as it plays a key role in how people interpret and understand a software system. Documentation quality becomes more important for AI models, where even with access to the underlying code, there is no guarantee if details such as, how was the data cleaned, the model's assumptions, or how the model works are available. Although standards for model documentation are a step towards addressing such underlying issues, without the corresponding ways of evaluating if that documentation is understandable, complete, and generally "good," we burden ourselves with bad documentation and all the problems that come along with it.

\begin{acks}
We would like to thank Karthik Muthuraman and Saishruthi Swaminathan for their helpful feedback and assistance in the evaluations.
\end{acks}

\bibliographystyle{ACM-Reference-Format}
\bibliography{main}

\end{document}